\documentclass[%
 reprint,
 amsmath,amssymb,
 aps,
]{revtex4-1}

\usepackage{graphicx}% Include figure files
\usepackage{dcolumn}% Align table columns on decimal point
\usepackage{bm}% bold math
\usepackage{color}

\begin{document}

\preprint{APS/123-QED}

\title{ Is Radiative Electroweak Symmetry Breaking Consistent with a 125 GeV Higgs Mass?}% Force line breaks with 
\author{T.G. Steele}
\author{Zhi-Wei Wang}%
\affiliation{%
Department of Physics and
Engineering Physics, University of Saskatchewan, Saskatoon, SK,
S7N 5E2, Canada
}%

\date{ arXiv submission 24 September 2012}% It is always \today, today,

\begin{abstract}
The mechanism of radiative electroweak symmetry breaking occurs through loop corrections, and unlike conventional symmetry breaking where the Higgs mass is a parameter, the radiatively-generated Higgs mass is dynamically predicted.
Pad\'e approximations and an averaging method 
are developed to extend the Higgs mass predictions in radiative electroweak symmetry breaking from five- to nine-loop order in the scalar sector of the Standard Model, resulting in an upper bound on the Higgs mass of $141\,{\rm GeV}$.   The mass predictions are well-described by a geometric series behaviour, converging to an asymptotic Higgs mass of 
$124\,{\rm GeV}$
  consistent with the recent ATLAS/CMS observations.  
  Similarly, we find that the Higgs self-coupling converges to  $\lambda=0.23$, which is significantly larger than its
  conventional symmetry breaking counterpart for a $124\,{\rm GeV}$  Higgs mass. In addition to this significant  enhancement of the  Higgs self-coupling and  $HH\rightarrow HH$ scattering, 
we find that 
Higgs decays to  gauge bosons are unaltered and the scattering processes  $W_{L}^{+}W_{L}^{+}\rightarrow HH$, $Z_{L}Z_{L}\rightarrow HH$ 
are also enhanced, providing signals to distinguish conventional and radiative electroweak symmetry breaking mechanisms. 
\end{abstract}

\maketitle

%\tableofcontents

%\section{\label{sec:level1}Introduction}

The observation of a $125\,{\rm GeV}$ Higgs candidate by ATLAS and CMS  \cite{:2012gk,:2012gu}, along with supporting evidence from CDF  and D0 \cite{Aaltonen:2012qt}, provides preliminary information for evaluating different mechanisms of electroweak (EW) symmetry breaking.  
Coleman \& Weinberg originally demonstrated that spontaneous symmetry breaking can occur through loop (radiative) corrections to the effective potential in the absence of a tree-level Lagrangian quadratic scalar term \cite{Coleman:1973jx}. This radiative EW symmetry-breaking mechanism is conceptually appealing because the Higgs mass is no longer a free-parameter as it is in conventional  EW symmetry breaking, but is a dynamical quantity which can be self-consistently predicted by the theory.  The absence of a conventional-symmetry-breaking quadratic scalar term also addresses aspects of the scale hierarchy problem \cite{Weinberg:1978ym} and the fine-tuning problem \cite{Susskind:1978ms}. 

The small-Higgs-coupling  radiative symmetry-breaking solution originally discovered  by Coleman \& Weinberg leads to an order $10\,{\rm GeV}$ Higgs mass which has been excluded by experiment. This result relies upon the dominance of gauge couplings over Yukawa couplings in the effective potential;  the large Yukawa coupling of the top quark (which was not known at the time of Coleman \& Weinberg) destabilizes the small-Higgs-coupling solution.  However, it has been demonstrated that a large-Higgs-coupling solution exists that results in a significantly larger Higgs mass prediction \cite{Elias:2003zm,Elias:2003xp}.  Similar radiative-symmetry-breaking solutions have been found in extensions of the Standard Model  \cite{Meissner:2006zh}.
%\footnote{Note that similar radiative-symmetry-breaking solutions have been found in extensions of the Standard Model  \cite{Meissner:2006zh}.}   

The large-Higgs-coupling solutions for radiative symmetry breaking are intrinsically challenging because higher-loop corrections can become important.   Fortunately, one can demonstrate that Yukawa and gauge couplings have minimal effect on the  analysis, and hence the scalar field sector of the Standard Model (a globally-symmetric $O(4)$ scalar field theory) captures the essential features of the effective potential and radiative symmetry breaking  in the full Standard Model \cite{Elias:2004bc}, providing a simpler field theory for evaluating higher-loop corrections.   In particular, at leading-order the largest secondary effect of the top-quark Yukawa coupling $x=x_t=0.025$ only has a 2\% effect  on the Higgs mass  \cite{Elias:2004bc,Chishtie:2010ni}.  
This arises as a combination of two main effects:  $x_t$ represents  15\% of the one-loop $\beta$ function for the Higgs self-coupling in the large-coupling solution,  and the $x$-independent tree-level contribution suppresses the $x$ dependence 
in the Higgs mass prediction.
%and the leading (tree-level) contribution to the Higgs mass $M_H$ is independent of $x$.  The resulting  Higgs mass in the large-coupling leading-logarithm solution has minimal dependence on the Yukawa coupling.  
%Although this justifies $O(4)$ scalar theory as a good approximation for radiative symmetry breaking, it is important to keep in mind that the $x_t$ is ultimately responsible for destabilizing any small coupling solutions emerging from the $O(4)$ scalar theory.  

Although such higher-loop calculations of the effective potential would initially seem daunting, 
 in the Coleman-Weinberg (CW) renormalization scheme \cite{Coleman:1973jx,Jackiw:1974cv}, the effective potential for scalar field theories with global $O(N)$ symmetry can be uniquely determined from the renormalization-group functions \cite{Chishtie:2007vd}.   Because the renormalization-group functions for $O(N)$-symmetric scalar $\phi^4$ theories are known to five-loop order in the MS scheme \cite{Kleinert:1991rg}, and methods are known for converting them to the CW scheme \cite{Ford:1991hw}, calculation of the five-loop effective potential in $O(N)$ $\phi^4$ theory has been achieved \cite{Chishtie:2007vd,Chishtie:2008va}.
The Higgs mass prediction resulting from these  higher-loop corrections shows evidence of slow convergence as loop order is increased,  resulting in  a Higgs mass  upper bound of $165\,{\rm GeV}$ from the five-loop effective potential  \cite{Chishtie:2010ni}.

The purpose of this paper is  to assess whether radiative EW symmetry-breaking can accommodate a $125\,{\rm GeV}$ Higgs mass as observed by ATLAS and CMS \cite{:2012gk,:2012gu} by estimating higher-loop effects on the Higgs mass prediction. 
This is achieved by exploiting the success of Pad\'e approximation methods for the renormalization-group functions of $O(N)$ $\phi^4$ theory \cite{Ellis:1996zn,Elias:1998bi,Chishtie:1998hs} to estimate  higher-loop contributions to the effective potential and Higgs mass.    We also argue that averaging subsequent orders of the effective potential further extends the estimates to two loops higher accuracy.    By combining these estimation methods, we obtain a nine-loop Higgs mass upper bound of $141\,{\rm GeV}$ and observe an empirical pattern of Higgs mass estimates that  extrapolates the Higgs mass prediction to a value in close agreement with ATLAS and CMS  \cite{:2012gk,:2012gu}.   However, mass predictions alone are not sufficient to distinguish conventional and radiative EW symmetry breaking.  Following Ref.~\cite{Chishtie:2005zi} we identify possible phenomenological signatures, including the Higgs self-coupling, that would distinguish a $125\,{\rm GeV}$ Higgs in conventional and radiative EW symmetry breaking.  

%\section{\label{sec:level2}Pad\'e Approximations for Effective Potential in  $O(N)$-Symmetric Massless $\lambda\phi^4$ Theory}

In $O(N)$-symmetric massless $\lambda\phi^4$ theory (i.e. the Standard Model scalar sector corresponds to $N=4$),
%\footnote{Note that the Standard Model scalar sector corresponds to $N=4$.} 
the effective potential in the CW scheme takes the form \cite{Coleman:1973jx}
\begin{align}
V\left(\lambda, \phi, \mu\right)=\sum_{n=0}^{\infty}\sum_{m=0}^{n}\lambda^{n+1}
T_{nm}L^m\phi^4
%&L=\log\left(\phi^2/\mu^2\right)\,,~\phi^2=\sum_{i=1}^N\phi_i^2\,,
%\nonumber
\end{align} 
where $L=\log\left(\phi^2/\mu^2\right)\,,~\phi^2=\sum_{i=1}^N\phi_i^2 $, and $\mu$ is the renormalization scale. The summation includes  leading logarithm (LL), next-to-leading logarithm (NLL), next-to-next-to-leading logarithm $N^2LL$, and in general $N^nLL$ terms.  The $N^nLL$ term $S_n$ can be isolated by rearranging the summation in the form 
\begin{align}
V\left(\lambda, \phi, \mu\right)=\sum_{n=0}^{\infty}\lambda^{n+1}S_n\left(\lambda L\right)\phi^4\label{expression of effective potential}
%S_n\left(\lambda L\right)&=\sum_{m=0}^{\infty}T_{n+m\, m}\left(\lambda L\right)^m\,.
%\nonumber
\end{align}
where $S_n\left(\lambda L\right)=\sum_{m=0}^{\infty}T_{n+m\, m}\left(\lambda L\right)^m$. The renormalization group (RG) equation
\begin{align}
&\left(\mu\frac{\partial}{\partial\mu}+\beta\left(\lambda\right)\frac{\partial}{\partial\lambda}
+\gamma\left(\lambda\right)\phi\frac{\partial}{\partial\phi}\right)V\left(\lambda, \phi, \mu\right)=0\label{renormalization group equation}\\
&\beta\left(\lambda\right)=\mu\frac{d\lambda}{d\mu}=\sum_{k=2}^{\infty}b_k\lambda^k\,,~
\gamma\left(\lambda\right)=\frac{\mu}{\phi}\frac{d\phi}{d\mu}=\sum_{k=1}^{\infty}g_k\lambda^k
\end{align}
leads to the following coupled differential equations for the functions $S_n(\xi)$  \cite{Chishtie:2007vd}
%\begin{align}
\begin{equation}
\begin{split}
0=&\left[\left(-2+b_2\xi\right)\frac{d}{d\xi}+\left(n+1\right)b_2+4g_1\right]S_n
%\nonumber
\\
&+\sum_{m=0}^{n-1}\left\{\left(2g_{n-m}+b_{n+2-m}\xi\frac{d}{d\xi}\right)\right.
%\nonumber
\\
&+\big[\left(m+1\right)b_{n+2-m}+4g_{n+1-m}\big]\bigg\}S_m\,,
\end{split}
\label{7}
\end{equation}
%\end{align}
where we show below that $g_1=0$.  We thus see that the $n+1$-loop renormalization-group functions are needed to determine $S_k$ for $k=0,1,2,\ldots , n$.  The boundary conditions $S_n\left(0\right)=T_{n0}$ needed to solve \eqref{7} emerge from the CW renormalization condition  
$\frac{d^4V}{d\phi^4}\vert_{\phi=\mu}=24\lambda$
\cite{Coleman:1973jx,Jackiw:1974cv}, 
 %\begin{equation}
%\left.\frac{d^4V}{d\phi^4}\right\vert_{\phi=\mu}=24\lambda\,,
%\label{8}
%\end{equation}
resulting in the constraints  \cite{Chishtie:2007vd}
%\begin{align}
\begin{equation}
\begin{split}
0&=16 \frac{d^4}{d\xi^4}S_k (0) + 80 \frac{d^3}{d\xi^3}S_{k+1}(0) + 140 \frac{d^2}{d\xi^2}S_{k+2}(0)
\\
&+ 100 \frac{d}{d\xi}S_{k+3} (0) + 24 S_{k+4} (0) ~~(k=0,1,2 \cdots)\,.
%\nonumber
\end{split}
\label{13}
\end{equation}
%\end{align}
The boundary condition $S_n(0)$ for the differential equation \eqref{7} can then be obtained by iteratively solving for the lower-order $S_k$, where $k=\{n-1,n-2,n-3,n-4\}$.  Thus in the CW scheme, the effective potential to $N^pLL$ order is uniquely determined by the $p+1$-loop RG functions.
However, since we only have the limited information of the renormalization group functions, we need to truncate the process at  a certain  $N^pLL$ order
%, which requires the introduction of a counter-term
%.  If we truncate the summation at $N^pLL$ order, we can write the effective potential at this order as
\begin{equation}
V_{p}=\sum_{n=0}^p\lambda^{n+1}S_n\left(\lambda L\right)\phi^4+\sum_{i=p+1}^{p+4}T_{i\,0}\lambda^{i+1}\phi^4\,,
\label{14}
\end{equation}
where the last term represents a  counter-term which is constrained by the CW renormalization condition. 
%\eqref{8}.  
It should be noted that this procedure can reproduce the explicit two-loop calculation of the effective potential \cite{Jackiw:1974cv}.

In general, the effective action also has divergences in the kinetic term 
%\begin{equation}
%\Gamma=\int d^4x\left[ -V(\phi)+\frac{1}{2}Z(\phi)\left(\partial_\mu \phi\right)^2+\ldots   \right] \,,
%\end{equation}
which are addressed in the CW scheme via a condition which maintains the tree-level form. 
%$Z(\phi)=1$.   
With this additional condition, the Higgs mass $M_H$ is given by 
\begin{equation}
M_H^2
=\frac{1}{Z}\frac{d^2V}{{d\phi}^2}\bigg|_{\phi=\mu}
=\frac{d^2V}{{d\phi}^2}\bigg|_{\phi=\mu}\,.
\label{mass prediction}
\end{equation}
where $Z(\phi)=1$ in the CW scheme.
Finally, the coupling $\lambda$ is determined by the spontaneous-symmetry-breaking condition that the effective potential has a non-trivial minimum
%\begin{equation}
$\frac{dV}{d\phi}\vert_{\phi=\mu}=0\,$.
%\label{coupling constant}
%\end{equation}
Contact with the Standard Model is achieved by identifying the scale $\mu$ with the EW scale $\mu=v=246.2\,{\rm GeV}$.
We note that although higher-loop calculations of the effective potential exist in other schemes, there are very few corresponding calculations of $Z(\phi)$, so for pragmatic purposes higher-loop calculations of the Higgs mass are currently  limited to the CW scheme.
By contrast, RG functions  $\tilde\beta$ and $\tilde\gamma$ are generally calculated in MS-like schemes, and hence it is necessary to convert these RG functions  to the CW scheme \cite{Ford:1991hw}.  %For our situation where there is a single coupling, the CW-scheme RG functions in \eqref{renormalization group equation} are related to their MS-scheme versions $\tilde\beta$ and $\tilde\gamma$   \cite{Ford:1991hw}.
%\begin{equation}
%\begin{split}
%\beta(\lambda)&=\frac{\tilde{\beta}(\lambda)}{1-\frac{\tilde{\beta}(\lambda)}{2\lambda}}\,,~
%\gamma(\lambda)=\frac{\tilde{\gamma}(\lambda)}{1-\frac{\tilde{\beta}(\lambda)}{2\lambda}}~,\\
%\tilde\beta\left(\lambda\right)=\mu\frac{d\lambda}{d\mu}&=\sum_{k=2}^{\infty}\tilde b_k\lambda^k\,,~
%\tilde\gamma\left(\lambda\right)=\frac{\mu}{\phi}\frac{d\phi}{d\mu}=\sum_{k=1}^{\infty}\tilde g_k\lambda^k~.
%\end{split}
%\label{MS-CW Transformation}
%\end{equation}
We can thus use the five-loop MS-scheme determinations of the $O(N)$-symmetric $\phi^4$ RG functions \cite{Kleinert:1991rg}
to determine their five-loop CW-scheme counterparts.  Note that in the MS-scheme $\tilde g_1=0$, and hence we also have $g_1=0$ in the CW scheme.

In Ref.~\cite{Chishtie:2010ni} it was demonstrated that for $p$ even, $V_p$ provides an upper bound on the Higgs mass $M_H$ which slowly drops from $221\,{\rm GeV}$ at one-loop order ($LL$ order) to $165\,{\rm GeV}$ at five-loop order ($N^4LL$).  Thus we can exclude radiative symmetry breaking if the upper bound drops below the ATLAS/CMS value of $125\,{\rm GeV}$.  We thus focus on improving the upper bound by approximating  higher-loop terms in the RG functions which enables higher-loop approximations to the effective potential.  

Pad\'e approximation methods, particularly when improved with an asymptotic error correction \cite{Ellis:1996zn,Samuel:1995jc,Ellis:1997sb}, have been successfully applied to the MS-scheme RG functions of $O(N)$ massive scalar field theory \cite{Ellis:1996zn,Elias:1998bi,Chishtie:1998hs}.  For example, using four-loop results as input, the Pad\'e-predicted and exact five loop term in the beta function agree to better than 5\% for $N=4$ \cite{Chishtie:1998hs}.  

For Pad\'e approximations to the MS-scheme beta function $\tilde\beta$ we write
\begin{equation}
\tilde\beta=\tilde b_2\lambda^2\left(1+\frac{\tilde b_3}{\tilde b_2}\lambda+ \frac{\tilde b_4}{\tilde b_2}\lambda^2
+\frac{\tilde b_5}{\tilde b_2}\lambda^3+\frac{\tilde b_6}{\tilde b_2}\lambda^4\right)
\end{equation}
and apply Pad\'e approximation methods to the bracketed quantity.  
Using the methods outlined in Refs.~\cite{Ellis:1996zn,Samuel:1995jc,Ellis:1997sb}, the asymptotic-improved  Pad\'e prediction of the $R_5x^5$ term with known coefficients $\{R_1,R_2,R_3,R_4\}$ in the series $P(x)=1+R_1x+R_2x^2+R_3 x^3+R_4 x^4$
%\begin{equation}
%P(x)=1+R_1x+R_2x^2+R_3 x^3+R_4 x^4
%\label{poly}
%\end{equation}
is given by
%\footnote{To our knowledge this result has not been presented elsewhere. }
\begin{equation}
R_5=\frac{R_4^2\left(R_1R_3^3-2 R_2^3R_4+R_1 R_2 R_3 R_4\right)}{R_3\left(2 R_1 R_3^3-R_2^2R_3^2-R_4 R_2^3\right)}~.
\end{equation}
From this expression, the resulting asymptotic-improved Pad\'e prediction of the $O(4)$ MS six-loop beta function is
$\tilde b_7=-1.07$.  Because we also want the seven-loop beta function, a $[2\vert 2]$ Pad\'e approximant to $P(x)$ is then used 
to predict its $x^5$ and $x^6$ terms.
%\begin{gather}
% S_{[2][2]}(x)=\frac{1+c_1x+c_2x^2}{1+d_1x+d_2x^2}\,,\label{22pade}\\
%d_1=\frac{R_1R_4-R_2R_3}{R_2^2-R_1R_3}\,,~
%d_2=\frac{R_3^2-R_2R_4}{R_2^2-R_1R_3}\,,~
%c_1=R_1+d_1\,,~
%c_2=R_2+d_1R_1+d_2\,.
%\label{c_2}
%\end{gather}
The resulting prediction of the six- and seven-loop MS-scheme $O(4)$ beta function coefficients are $\tilde b_7=-0.992$ and $\tilde b_8=1.96$.  The agreement of the $[2\vert2]$ and asymptotic-improved predictions for $\tilde b_7$  to approximately 10\% provides a validation of the $[2\vert2]$ methodology and establishes a characteristic error for our analysis below.   Converting these MS-scheme beta function coefficients to the CW scheme results in $b_7=-0.695$ and $b_8=1.37$.

For Pad\'e approximations to the MS-scheme  anomalous dimension we follow the same procedure except
%we write
%\begin{equation}
%\tilde \gamma=\tilde g_2\lambda^2\left(1+\frac{\tilde g_3}{\tilde g_2}\lambda+ \frac{\tilde g_4}{\tilde g_2}\lambda^2
%+\frac{\tilde g_5}{\tilde g_2}\lambda^3\right)~.
%\end{equation} 
%Note that 
there is less information than in the beta function because the leading-order $\tilde g_1$ term in the anomalous dimension is zero. Given knowledge of  $\{R_1,R_2, R_3\}$ in the series $P(x)$, the asymptotic-improved Pad\'e prediction of $R_4$ \cite{Elias:1998bi}
%\begin{equation}
%R_4=\frac{R_3^2\left(R_2^3+R_1R_2R_3-2R_1^3R_3\right)}{R_2\left(2R_2^3-R_1^3R_3-R_1^2R_2^2\right)}\,,
%\label{R_4}
%\end{equation} 
results in the asymptotic-improved Pad\'e prediction of the six-loop coefficient $\tilde g_6=-0.692\times 10^{-3}$ and $g_6=-0.135\times 10^{-3}$ after conversion to the CW scheme.   As argued above, we can now use  $\tilde  g_6$ to form a $[2\vert 2]$ Pad\'e  approximant to predict the seven-loop MS coefficient $\tilde g_7=0.961\times 10^{-3}$ which corresponds to CW-scheme value $g_7=-0.225\times 10^{-4}$.

Equipped with Pad\'e estimates of the RG functions up to seven-loop order, we can solve \eqref{7} and \eqref{13} to obtain $S_5$ and $S_6$, which enables the construction of the $N^6LL$ effective potential $V_6$.   Analysis of the effective potential  results in the Higgs mass $M_H=150\,{\rm GeV}$ and the CW-scheme weak-scale coupling $\lambda(v)=0.308$. Including $10\%$ uncertainties in the RG coefficients only gives a  $0.1\,{\rm GeV}$ Higgs mass difference which shows the method is quite robust.

\begin{figure}[htb]
\centering
\includegraphics[scale=1]{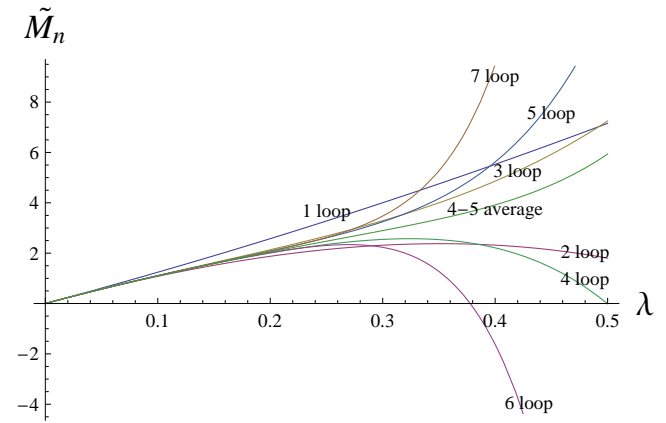}
\caption{The dimensionless quantity 
%$\tilde M_n=\left.\frac{1}{v^2}\frac{d^2\left(V_n-K_n\phi^4\right)}{d\phi^2}\right\vert_{\phi=v}$ 
$\tilde M_n$ \eqref{tilde_M}
is plotted as a function of $\lambda$ for the $O(4)$ scalar theory. Upper curves represent the  even $N^pLL$ ($p+1$-loop) orders ($p=0,2,4,6$) and the lower curves represent the 
%$N^pLL$ ($p+1$-loop) 
 odd orders ($p=1,3,5$).  The average of the four- and five-loop contributions is also shown. } 
\label{fig3}
\end{figure}

%\section{\label{sec:level3}Higgs Mass Extrapolation to Higher-Loop Order}
Now, we develop methods to extrapolate the Higgs mass estimates to higher-loop orders.  We begin with the averaging method
motivated by the field-theoretical contributions to the Higgs mass 
 in the absence of a counter-term contribution 
 \begin{equation}
 \tilde{M}_n=\frac{1}{v^2}\frac{d^2\left(V_n-K_n\phi^4\right)}{d\phi^2}\bigg|_{\phi=\mu}\,,
 \label{tilde_M}
  \end{equation}
where $K_n=\sum_{i=n+1}^{n+4}T_{i,0}\lambda^{i+1}$ is the corresponding counter-term at that order. 
The quantity $\tilde M_n$ is shown in Fig.~\ref{fig3}, and
as argued in Ref.~\cite{Chishtie:2010ni},   demonstrates that the effective potential over-estimates the Higgs mass at even orders and under-estimates it at odd orders.   Moreover, we can imagine that at higher orders, the field theoretical contributions to the Higgs mass  will lie in the envelope between the even and odd orders.  Indeed, for  small $\lambda$ Fig.~\ref{fig3} already shows close agreement between  even and odd orders.  It thus seems plausible that the average of an even- and odd-order effective potential will provide a better  approximation to the full effective potential than a single-order result.  
This averaging method can be justified by identifying $\tilde M_n$ as the Eq.~\eqref{14} partial sum of  $S_n$ contributions that alternate in sign as demonstrated by  Fig.~\ref{fig3}.  
In particular,  moving from $p$-loop order to $p+1$-loop order involves the addition of $S_p$, and the sign of this contribution sequentially raises/lowers the curves in  Fig.~\ref{fig3}. 
Thus for any fixed value of $\lambda$, we can employ Euler's transformation for alternating series to accelerate its convergence \cite{numerical_recipes}:
\begin{equation}
\sum_{n=0}^{\infty}(-1)^nu_n=u_0-u_1
%+u_2\cdots
+\ldots
-u_{N-1}+\sum_{s=0}^{\infty}\frac{(-1)^s}{2^{s+1}}\left[\Delta^su_N\right]
\end{equation}
and setting $s=0$ as the lowest order approximation, we obtain:
%\begin{equation}
%\sum_{n=0}^{\infty}(-1)^nu_n\approx u_0-u_1+\ldots-u_{N-1}+\frac{1}{2}u_N~.
%\end{equation}
% We can express this approximation as the average of subsequent partial sums  $P_N=\sum_{n=0}^{N}(-1)^nu_n$,
\begin{equation}
%\begin{split}
\sum_{n=0}^{\infty}(-1)^nu_n
%&
\approx u_0-u_1+\ldots-u_{N-1}+\frac{1}{2}u_N =\bar P_N \,,
%\\
%&=\frac{P_N+P_{N-1}}{2}\equiv \bar P_N ~.
%\end{split}
\end{equation}
where the  partial sums  $P_N=\sum_{n=0}^{N}(-1)^nu_n$ and $\bar P_N=\frac{1}{2}\left(P_N+P_{N-1}\right)$.
In our case, the averaged effective potential $\bar V_n$ can be written as $\bar V_n=\frac{1}{2}\left(V_n+V_{n-1}\right)$.
%\begin{equation}
%\begin{split}
%\bar V_n&=\frac{V_n+V_{n-1}}{2}
%\\&=
%\left(\lambda S_0+\lambda^2 S_1+\cdots+\lambda^nS_{n-1}+\frac{1}{2}\lambda^{n+1}S_n+
%%\sum_{i=n+1}^{n+4}T_{i0}\lambda^{i+1}
%\bar K_n
%\right)\phi^4
%\end{split}
%\end{equation}
%where $\bar K_n=K_n/2+K_{n-1}/2$ represents the averaged  counter-terms from Eqs.~\eqref{14}, \eqref{tilde_M}.

For example, the average of four-loop ($N^3LL$) and five-loop ($N^4LL$) contributions to the effective potential  
$\bar V_4=V_3/2+V_4/2=\left(\lambda S_0+\ldots+\lambda^4S_{3}+\frac{1}{2}\lambda^{5}S_4+\bar K_4\right)\phi^4$
%$\bar V_4=V_3/2+V_4/2$.
%\begin{equation}
%\bar V_4=\frac{1}{2}V_4+\frac{1}{2}V_5\,.
%\end{equation}
 shown in Fig.~\ref{fig3}
%Using the renormalization condition Eq.~\eqref{8}, we can determine all the counter-terms appropriate to the average, and then find 
gives the Higgs mass prediction $M_H=153\,{\rm GeV}$ and the corresponding coupling $\lambda=0.418$.  Although the averaging results in a coupling close to the three-loop value, the mass is in close agreement with the seven-loop Pad\'e result.    
%The behaviour of  the coupling can be visualized by 
% a horizontal line, corresponding to the same field-theoretical contribution to the Higgs mass, which then intersects the four- and five-loop average at a larger coupling than the seven-loop Pad\'e result, precisely as 
 %%emerges from the detailed analysis 
 %presented in Table~\ref{res_tab}.
 Thus the average of the five-loop effective potential with its lower-loop 
four-loop counterpart, leads to a much better Higgs mass estimate than the five-loop contribution alone.

The same pattern holds at lower orders as well; the Higgs mass and coupling resulting from averaging  the two- and three-loop effective potentials is in remarkably-close agreement with the five-loop Higgs mass  and one-loop coupling (see Table~\ref{res_tab}).  Based on this pattern that the average of the $n-1$ and $n$-loop effective potentials approximates the $n-2$ loop coupling and $n+2$ loop Higgs mass, we expect that the average of the six- and seven-loop Pad\'e approximations to the effective potential will provide a good estimate of the nine-loop Higgs mass prediction and the five-loop coupling.  Using this method, our nine-loop estimates are $M_H=141\,{\rm GeV}$ with the corresponding coupling  
 $\lambda(v)=0.352$.  We note that the agreement between the five-loop  coupling and the six- and seven-loop average (see Table~\ref{res_tab}) provides further confirmation of the pattern, and gives us confidence in the nine-loop Higgs mass estimate.

\begin{table}[ht]
\centering
  \begin{tabular}{|| l | l | l | l || l | l | l | l ||}
    	\hline
Loop & $\lambda$ & $M_{H}$ & $\lambda_{CSB}$ & Average & $\lambda$ & $M_{H}$ & $\lambda_{CSB}$ \\ \hline $1$ loop & $0.534$ & $221$ & $0.101$ & & & & \\ \hline $3$ loops & $0.417$ & $186$ & $0.072$ &
2,3 loop & $0.514$ & $167$ & $0.230$ \\ \hline $5$ loops & $0.354$ & $165$ & $0.056$ & 4,5 loop & $0.418$ & $153$ & $0.194$ \\ \hline $7$ loops & $0.308$ & $150$ & $0.047$ & 6,7  loop & $0.352$ & $141$ & $0.041$ \\ \hline
  Extrapolate & $0.233$ & $124$ & $0.032$ &  &  &  & \\ \hline
\end{tabular}
\caption{Higgs mass in GeV and self-coupling predictions at  different loop orders in both the standard (left half) and averaging method (right half).  The extrapolated values emerging from the geometric series behaviour are also shown. For comparison, 
the Higgs coupling $\lambda_{CSB}$ in conventional symmetry breaking corresponding to the predicted Higgs mass is also provided.}
\label{res_tab}
\end{table}

\begin{figure}[htb]
\centering
\includegraphics[scale=1]{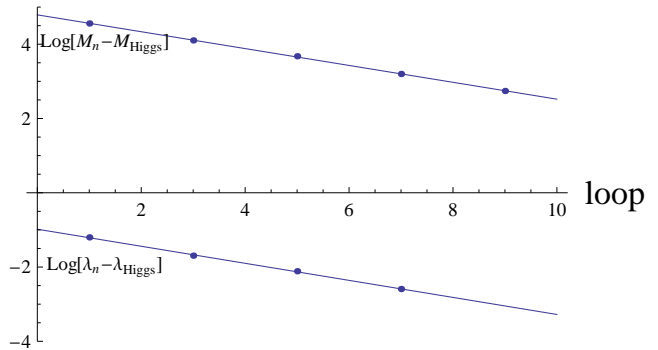}
\caption{The  quantities $\log\left(M_n-M_{Higgs}\right)$ and $\log\left(\lambda_n-\lambda_{Higgs}\right)$  are plotted versus loop order $n$ for Table~\ref{res_tab} values with $M_{Higgs}=125\,{\rm GeV}$ and  $\lambda_{Higgs}=0.23$. The  lines are a  linear fit to the data points based on the geometric series \eqref{geometric}.
}
\label{linear_fig}
\end{figure}

\begin{figure}[htb]
\centering
\includegraphics[scale=1]{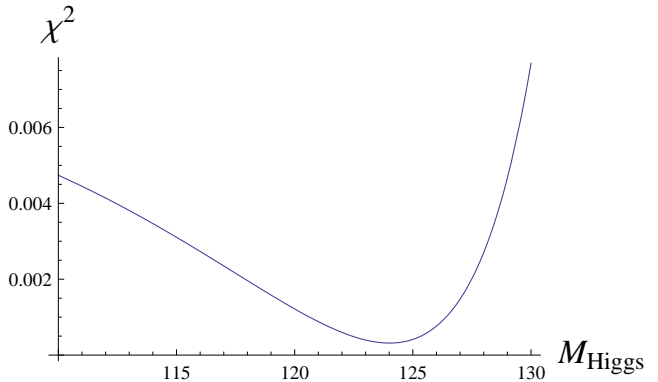}
\caption{
The $\chi^2$ of the linear fit of Table~\ref{res_tab} values to $\log\left(M_n-M_{Higgs}\right)$ is plotted as a function of $M_{Higgs}$ resulting in the least-squares prediction  $M_{Higgs}=124\,{\rm GeV}$.  
}
\label{chi2fig}
\end{figure}

Thus by combining Pad\'e estimates and the averaging method, the seven- and nine-loop Higgs mass predictions have been estimated.  These estimates demonstrate a continued slow convergence towards a Higgs mass bounded from above by $141\,{\rm GeV}$.   To determine whether the Higgs mass will eventually converge to a value consistent with the $125\,{\rm GeV}$ Higgs mass seen by ATLAS/CMS \cite{:2012gk,:2012gu}, we first note that the differences between subsequent loop orders in Table~\ref{res_tab} decrease in a fashion consistent with a geometric series
\begin{equation}
M_n-M_{Higgs}=\Lambda \sigma^n~,
\label{geometric}
\end{equation}
where $M_n$ is the $n$-loop Higgs prediction, $\Lambda$ has dimensions of mass, and $\sigma<1$ is a dimensionless quantity which leads to $\lim_{n\to\infty}M_n=M_{Higgs}$.  
In Figure~\ref{linear_fig} the plot of $\log\left(M_n-M_{Higgs}\right)$ shows clear linear behaviour with $n$ consistent with the geometric series \eqref{geometric} when $M_{Higgs}=125\,{\rm GeV}$.  The dependence of the $\chi^2$ deviation from this linear fit is shown as a function of $M_{Higgs}$ in Fig.~\ref{chi2fig}, providing an optimized value $M_{Higgs}=124\,{\rm GeV}$. 
We thus speculate that the radiatively-generated Higgs mass ultimately converges
to a value consistent with the $125\,{\rm GeV}$ ATLAS/CMS value \cite{:2012gk,:2012gu}.
A similar geometric series pattern exists for the Higgs self-coupling;  Figure~\ref{linear_fig} shows linear behaviour for  $\log\left(\lambda_n-\lambda_{Higgs}\right)$ for the least-squares optimized value  $\lambda_{Higgs}=0.23$.
The similarity in slope between the mass and coupling plots in Fig.~\ref{linear_fig} is intriguing; we speculate that this is connected to the underlying rate of convergence of the effective potential.

The extrapolated value $\lambda_{Higgs}$ is a factor of 2 smaller than the leading logarithm result, so it is necessary to re-examine whether the scalar field theory sector of the Standard Model is still a valid approximation.   Because the top-quark Yukawa coupling %$x_t=g_t^2/4\pi^2=0.025$ 
is the dominant secondary effect on the Higgs mass \cite{Elias:2004bc}, 
%in Fig.~\ref{yukawa_fig} we 
we have calculated the leading-logarithm (one-loop) ratio of the Higgs mass with top-quark effects included and omitted.  For the extrapolated value $\lambda_{Higgs}=0.23$, the top-quark effects are less than a 5\% effect at leading-log order and hence the scalar field sector still captures the dominant features of the Standard Model.  

%\begin{figure}[htb]
%\centering
%\includegraphics[scale=1]{Higgs_mass_yukawa.eps}
%\caption{
%The ratio of the Higgs mass with the physical top-quark Yukawa coupling $x_t=0.025$ and with top-quark coupling omitted ($x_t=0$) is shown as a function of the Higgs self-coupling.  The shown range of $\lambda$ contains the full range of Table~\ref{res_tab}. 
%}
%\label{yukawa_fig}
%\end{figure} 

%\section{\label{sec:level4}Discussion}
Using Pad\'e approximation methods and an averaging technique, we have extended the radiatively-generated Higgs mass prediction to nine-loop order.  Two important trends emerge from this result: both the Higgs mass and CW-scheme coupling $\lambda$ decrease with increasing loop order. 
 Both the Higgs mass and self-coupling  are well-described by a geometric series in the loop-order, converging to approximately $M_{Higgs}=124\,{\rm GeV}$ and $\lambda_{Higgs}=0.23$. 
 %For this coupling, the curves in Fig.~\ref{fig3} begin to coincide.
% We observe an interesting combination of effects: the overlapping region of the even- and odd-order curves of Fig.~\ref{fig3} extends to larger $\lambda$ with  increasing loop level and the predicted $\lambda$ value decreases. The combination of these two effects  drives  $\lambda$  to  a value where both the even- and odd-order curves in Fig.~\ref{fig3} give the same mass contribution,  indicating overall consistency between the various loop-order contributions to the Higgs mass.

The value of the coupling  provides  a phenomenological signal that distinguishes between radiative and conventional EW symmetry breaking.  For example, in Table~\ref{res_tab} the conventional symmetry-breaking  $\lambda$ value corresponding to the predicted Higgs mass is smaller than the radiatively-generated value at all orders, implying a significant enhancement of Higgs-Higgs scattering in radiative EW symmetry-breaking \cite{Elias:2003zm}.   This trend
 is upheld in the extrapolation to  $\lambda_{Higgs}=0.23$ and a $125\,{\rm GeV}$ Higgs mass.  
It seems feasible for the Higgs self-coupling to be measured by the LHC \cite{Papaefstathiou,Dolan:2012rv}; an enhancement  of the coupling compared to conventional symmetry breaking  could be evidence for the radiative scenario.

However,  the Goldstone-boson replacement theorem  \cite{Lee:1977eg}
%(which approximates high-energy gauge-boson processes by their Goldstone boson equivalent via the effective potential \cite{Lee:1977eg})
 leads to identical results for the Higgs decay processes  $H\rightarrow W_{L}^{+}W_{L}^{-}$, $H\rightarrow Z_{L}Z_{L}$ in conventional and radiative symmetry breaking  independent of the extrapolated Higgs coupling, and similar equivalences are found for the scattering processes $W_{L}^{+}W_{L}^{-}\rightarrow W_{L}^{+}W_{L}^{-}$, $W_{L}^{+}W_{L}^{-}\rightarrow Z_{L}Z_{L}$ \cite{Chishtie:2005zi}.  This implies that these Higgs decays and gauge boson scattering processes are unable to distinguish between radiative and conventional EW symmetry breaking.  By contrast,  the  processes $W_{L}^{+}W_{L}^{+}\rightarrow HH$, $Z_{L}Z_{L}\rightarrow HH$ are enhanced in radiative EW symmetry breaking independent of the extrapolated Higgs coupling.   For example,  the seven-loop Pad\'e prediction  leads to a three-fold  enhancement comparable to the lower-loop analysis of Ref.~\cite{Chishtie:2005zi}.

In summary, we have combined Pad\'e approximation methods with averaging techniques to extend Higgs mass predictions in radiative EW symmetry breaking to the nine-loop  estimate $M_H=141\,{\rm GeV}$ for the upper bound on the Higgs mass.  Evidence of  geometric-series convergence of the Higgs mass to $125\,{\rm GeV}$ suggests that radiative EW symmetry breaking is a viable mechanism for the ATLAS/CMS observation of a Higgs.    Similar evidence of geometric series convergence for the Higgs self-coupling leads to the corresponding limiting value of the radiatively-generated Higgs coupling $\lambda_{Higgs}=0.23$,  a value significantly larger than its conventional symmetry-breaking counterpart.
This implies that
the processes $HH\to HH$,  $W_{L}^{+}W_{L}^{+}\rightarrow HH$, $Z_{L}Z_{L}\rightarrow HH$ are enhanced, so  
discrepancies between experiment and conventional symmetry-breaking  predictions may provide signals of  radiative symmetry breaking.

\noindent
{\bf Acknowledgements:}  TGS is grateful for financial support from the Natural Sciences and Engineering Research Council of Canada (NSERC).

\end{document}